\documentclass[12pt]{article}
\usepackage{epsf}
\begin{document}
\begin{titlepage}
\title{Tree and penguin amplitudes from $B \to \pi\pi, K\pi, K\bar{K}$}
\author{
{Piotr \.Zenczykowski }\\
{\em Division of Theoretical Physics},
{\em Institute of Nuclear Physics,}\\
{\em Polish Academy of Sciences,}\\
{\em Radzikowskiego 152,
31-342 Krak\'ow, Poland}\\
E-mail: piotr.zenczykowski@.ifj.edu.pl
}
\maketitle
\begin{abstract}
The question of the relative size of tree and penguin amplitudes 
is analyzed using the data on $B \to \pi\pi$, $B^+ \to \pi^+K^0$,
and $B^+ \to K^+\bar{K}^0$ decays. Our discussion involves an estimate of $SU(3)$
breaking in the final quark-pair-creating hadronization process.
The estimate is based on Regge phenomenology, which many years ago proved 
very successful in the description of soft hadronic physics.
Accepting the Regge prediction as solid, it is then shown that the relative size and phase of the two parts of the
penguin amplitude can be unambiguously
extracted from the data on the decays considered. 
This enables fixing the $C/T$ ratio of `true' tree amplitudes, which -
on the basis of the existing data - is shown to
be small (of the order of 0.2).

\end{abstract}
PACS: 13.25.Hw; 12.15.Ji; 12.40.Nn; 14.40.Nd 
\vfill
\end{titlepage}

\section{Introduction}
Rare charmless nonleptonic B meson decays provide us with a lot of information
 on weak interactions of quarks and the Cabibbo-Kobayashi-Maskawa (CKM) matrix. 
 Yet, since quarks are forever confined, 
 this information is necessarily blended with strong interaction effects. 
 Additionally, there may be contributions from New Physics beyond the Standard Model. 
Since this New Physics is by definition unknown, 
proper disentangling of all three effects requires a thorough understanding of the 
strong interaction part. 

Unfortunately, the calculation of low-energy quark-confining
strong interaction effects from first principles is
at present impossible.
Consequently, the only way to achieve a reliable understanding of these effects 
is through their
parametrization and 
subsequent extraction of relevant parameters from the experimental data.
For example, simple quark-level arguments suggest that the ratio $C/T$
of the so-called colour-suppressed ($C$) and tree ($T$) diagrammatic 
amplitudes (see eg. \cite{GHLR}) should be small.
Yet, since the time of the theoretically clean analysis of ref.
\cite{Buras2004}, it is known that the effective tree and colour-suppressed
amplitudes in $B\to \pi\pi$ decays are roughly equal in absolute
magnitudes, thus contradicting simple quark-level expectations.
Hence, additional quark-level \cite{Beneke,KouPham} and/or hadron-level \cite{Zen2004} effects 
must be important.

In fact, as various authors have argued, the effective tree ($\tilde{T}$) 
and colour-suppressed ($\tilde{C}$)
amplitudes involve contributions from penguin diagrams as well
(see eg. \cite{Buras2004,Chiang2004}).
Disentangling this `second' penguin contribution from the `true' $C$ and $T$ amplitudes
is not possible on the basis of $B\to \pi\pi$ data alone.
For this reason, ref.\cite{SZ2005} considered $B \to \pi \pi$ decays in
conjunction with $B\to \pi \rho$ and $B \to \pi
\omega$ processes. Since under very reasonable assumptions it may be shown
 that both $B \to \pi\pi$ and $B \to \pi \rho, \pi \omega$ processes depend on
 the {\it same} {\it ratio} of $C/T$, 
 the number of equations constraining $C/T$ increases, thereby enabling
  extraction of the latter from the data. 
 In ref. \cite{SZ2005} it was then shown that the data allow a
 solution for $C/T$ which is in agreement with the simple 
 quark-level expectations.
 
 Alternatively, one may supplement the data on $B\to \pi\pi$ decays with those
 on $B^+ \to K^+\bar{K}^0$. This was the route attempted in ref. \cite{Baek2008}. 
 The problem with that route is that one has to know how to treat $SU(3)$
 breaking.
 
In this paper, we analyse $B \to \pi\pi$ and $B^+ \to K^+\bar{K}^0$ decays in
conjunction with $B^+ \to \pi^+ K^0$ processes, and under a specific assumption
concerning the pattern of $SU(3)$ breaking. 
The assumption on $SU(3)$ breaking
adopted by us has been
experimentally known to be correct since the mid-seventies of the last century.
We show that the data on $B\to \pi\pi, K^+\bar{K}^0,\pi^+K^0$ point then
unambiguously
 toward a small $|C/T|$ ratio, of the order of $0.2$.

The paper is organized as follows. In Section 2 our main
definitions and conventions
are set and the basic analysis of
the $B\to \pi\pi$ sector \cite{Buras2004,SZ2005}
is repeated with the new,
more precise data.
Section 3 contains the analysis of the $B^+ \to \pi^+ K^0, K^+\bar{K}^0$ decays.
This involves a Regge estimate of $SU(3)$ breaking in the hadronization stage,
which is crucial in bringing the extracted value of $|C/T|$ into agreement with
the expectations. 
We conclude in Section 4. 

\section{Decays $B \to \pi \pi$}

Below we employ the notation used in ref. \cite{SZ2005}.
The diagrammatic approach gives the following expressions for the
amplitudes in the $B \to \pi \pi$ decays:
\begin{eqnarray}
-\sqrt{2}A(B^+ \to \pi^+\pi^0)&=&\tilde{T}+\tilde{C},\nonumber\\
-A(B^0_d\to\pi^+\pi^-)&=&\tilde{T}+P_c,\nonumber\\
\label{Btopipi}
-\sqrt{2}A(B^0_d\to \pi^0\pi^0)&=&\tilde{C}-P_c,
\end{eqnarray}
where we have kept the leading terms only, i.e. the contributions from 
effective tree $\tilde{T}$,
colour-suppressed tree $\tilde{C}$, and penguin amplitudes $P_c$.
The effective amplitudes present in Eq. (\ref{Btopipi})
are related  to `true' tree $T$ and  colour-suppressed $C$ amplitudes, 
as well as to two contributions to the total penguin amplitude
$P$:
\begin{equation}
\label{Ppipi}
P=-\lambda_u^{(d)}{\cal{P}}_{tu}-\lambda_c^{(d)}{\cal{P}}_{tc},
\end{equation}
where
\begin{eqnarray}
\lambda_q^{(k)}&=&V_{qk}V^*_{qb},\nonumber\\
{\cal{P}}_{tq}&=&{\cal{P}}_t-{\cal{P}}_q,
\end{eqnarray}
with 
$V_{qk}$ being the CKM elements, and
${\cal{P}}_q$ being the contribution from quark $q$ running inside the
penguin loop.

After introducing
\begin{eqnarray}
\label{Pqdef}
P_q\equiv -\lambda_c^{(d)}{\cal{P}}_{tq}=A 
\lambda^3 {\cal{P}}_{tq},
\end{eqnarray}
where $A$ and $\lambda$ are Wolfenstein parameters,
the relevant formulas for the effective amplitudes are:
\begin{eqnarray}
P_c&=&A\lambda^3{\cal{P}}_{tc},\nonumber\\
\tilde{T}&=&e^{i\gamma}(T-R_bP_u),\nonumber\\
\tilde{C}&=&e^{i\gamma}(C+R_bP_u),
\end{eqnarray}
where  the weak phase factor has been explicitly
factored out from the `true' tree amplitudes $C$, $T$,
and $R_b=\sqrt{\bar{\rho}^2+\bar{\eta}^2}=0.37 \pm 0.02$ (all experimental numbers are 
taken from HFAG \cite{HFAG}).

Following refs \cite{Buras2004},\cite{SZ2005},
we define
\begin{eqnarray}
d e^{i\theta}&=&-e^{i\gamma}\frac{P_c}{\tilde{T}}=
\frac{P_c}{R_bP_u-T},\nonumber\\
\label{xdef}
x e^{i\Delta}&=&\frac{\tilde{C}}{\tilde{T}}=\frac{C+R_bP_u}{T-R_bP_u}.
\end{eqnarray}
Then, asymmetries $A^{dir}_{\pi^+\pi^-}$, $A^{mix}_{\pi^+\pi^-}$, 
with
experimental values of
\begin{eqnarray}
A^{dir}_{\pi^+\pi^-}&=&C_{\pi\pi}=-0.38\pm 0.06,\nonumber\\
A^{mix}_{\pi^+\pi^-}&=&-S_{\pi\pi}=+0.65\pm 0.07,
\end{eqnarray}
are expressed in terms of $d$ and $\theta$ as
\begin{eqnarray}
A^{dir}_{\pi^+\pi^-}&=&-\frac{2 d \sin \theta \sin \gamma}
{1-2 d \cos \theta \cos \gamma + d^2},\nonumber\\
A^{mix}_{\pi^+\pi^-}&=&\frac{\sin (2\beta + 2 \gamma)-2 d \cos \theta \sin
(2 \beta + \gamma)+d^2 \sin (2 \beta)}{1-2 d \cos \theta \cos \gamma +d^2}.
\end{eqnarray}
Using $\beta = 21.2^o$, $\gamma = 65.5^o $ 
(as in ref. \cite{Baek2008}), one then solves the above equations for $d$
and $\theta$ to find:
\begin{eqnarray}
d&=&0.51^{+0.10}_{-0.08},\nonumber\\
\label{dtheta}
\theta&=&140^o \pm 6^o.
\end{eqnarray}
When the following CP-averaged $B \to \pi\pi$ branching ratios 
$R^{\pi\pi}_{+-}$, $R^{\pi\pi}_{00}$ are defined:
\begin{eqnarray}
R^{\pi\pi}_{+-}&\equiv &2\frac{\langle{\cal{B}}(B^{+}\to\pi^{+}\pi^0)
\rangle_{CP}}
{\langle{\cal{B}}(B_d\to\pi^+\pi^-)\rangle_{CP}}\frac{\tau_{B^0_d}}{\tau_{B^+}},\nonumber\\
R^{\pi\pi}_{00}&\equiv &2\frac{\langle{\cal{B}}(B_d\to \pi^0\pi^0)\rangle_{CP}}
{\langle{\cal{B}}(B_d\to\pi^+\pi^-)\rangle_{CP}},
\end{eqnarray}
they may be expressed in terms of the yet undetermined parameters $x$, $\Delta$
as
\begin{eqnarray}
R^{\pi\pi}_{+-}&=&
\frac{1+2x \cos \Delta+x^2}{1-2 d \cos \theta \cos \gamma + d^2},\nonumber\\
\label{Rpipis}
R^{\pi\pi}_{00}&=&
\frac{d^2+2 d x \cos (\Delta-\theta) \cos \gamma + x^2}
{1-2 d \cos \theta \cos \gamma +d^2}.
\end{eqnarray}
For the values $\frac{\tau_{B^+}}{\tau_{B^0_d}}=1.073\pm 0.008$,
$\langle{\cal{B}}(B^{\pm}\to\pi^{\pm}\pi^0)\rangle_{CP}=5.59^{+0.41}_{-0.40}$,
$\langle{\cal{B}}(B_d\to\pi^+\pi^-)\rangle_{CP}=5.16\pm0.22$, and
$\langle{\cal{B}}(B_d\to \pi^0\pi^0)\rangle_{CP}=1.55\pm0.19$ (in units of $10^{-6}$)
one finds
\begin{eqnarray}
R^{\pi\pi}_{+-}&=&2.02 \pm 0.17,\nonumber\\
R^{\pi\pi}_{00}&=&0.60 \pm 0.08.
\end{eqnarray}
Using the central values for $d$ and $\theta$, 
Eqs (\ref{Rpipis}) then yield
\begin{eqnarray}
x&=&1.06^{+0.06}_{-0.07},\nonumber\\
\label{xDelta}
\Delta&=&-{59^o}^{+12^0}_{-11^o}.
\end{eqnarray}
A large (close to $1$) value of $x$ constitutes a problem in those approaches 
in which the contribution of $P_u$ is neglected
 since  for
small $P_u$  we have $|C/T| \approx x $ (Eq. (\ref{xdef})).

If $P_u$ is not neglected,  then --  as proposed in ref. \cite{SZ2005} --
using information 
on $B \to \pi \rho$ and $B \to \pi \omega$ decays and
making a very reasonable physical assumption concerning the creation of
the $q\bar{q}$ pair in the final hadronization process,
one can try to extract the true ratio $C/T$.
 In fact, it was shown in ref. \cite{SZ2005} 
 that there exists a solution with a small value of $|C/T|$ (of the order of
0.3).
Extraction of $C/T$ is thus possible with the help of
 additional
information available from other data.

The method of ref. \cite{Baek2008} uses for that purpose
the data on $B \to K\bar{K}$. Under the
assumption of exact $SU(3)$ it shows that the true $C/T$ may indeed be small.
However, when a specific way of $SU(3)$ breaking is considered, ref.\cite{Baek2008}
predicts significantly increased central values of $C/T$ (albeit the
corresponding errors also increase).

\section{Decays $B \to \pi K, K\bar{K}$}
The problem with the method of ref. \cite{Baek2008} is that 
it assumes either
exact $SU(3)$, or - as we will show - an inadequate estimate 
of $SU(3)$ breaking.
In the following, we show how $SU(3)$ breaking in the hadronization stage 
should be included,
and how this affects the discussion of the tree and penguin amplitudes (other
studies of $B \to K\bar{K}$ decays may be found in refs 
\cite{Recksiegel,He}).
In order to present the problem with $SU(3)$ breaking clearly, 
it is appropriate 
to consider 
two different
but closely  related pure
penguin processes, i.e. the decays $B^+ \to K^0\pi^+$ and $B^+ \to K^+\bar{K^0}$, and to
discuss them in conjunction.

\subsection{$B^+ \to \pi ^+K^0$}
When compared with the $B \to \pi\pi$ amplitudes,
the $B^+ \to \pi^+ K^0$ amplitude differs in that it is
now an
$s$ quark and not a $u$ or  $d$ quark that is being produced.
The relevant difference enters through the CKM factors $\lambda^{(k)}_q$ only. 
When $\lambda^{(k)}_q$'s are
factored out, the remaining factors in penguin amplitudes 
(i.e. ${\cal{P}}_{tu}$, ${\cal{P}}_{tc}$) should be the same in both
$B \to \pi\pi$ and $B\to \pi K$
(cf. \cite{SZ2005}). Thus, the $B^+ \to \pi^+ K^0$ amplitude 
is given
by
\begin{eqnarray}
A(B^+\to\pi^+K^0)&=&P'=-\lambda^{(s)}_u{\cal{P}}_{tu}-\lambda^{(s)}_c
{\cal{P}}_{tc}.
\end{eqnarray}
When 
the ratio of ${\cal{P}}_{tu}/{\cal{P}}_{tc}$
is expressed in terms of $P_u/P_c$, one obtains
\begin{eqnarray}
\label{Pprim}
P'&=&-\,\frac{1}{\sqrt{\epsilon}}\,P_c \,(1+\epsilon R_b 
\frac{P_u}{P_c}\,e^{i \gamma}),
\end{eqnarray}
where 
\begin{equation}
\epsilon = \frac{\lambda^2}{1-\lambda^2}\approx 0.05,
\end{equation}
and the factor of $1/\sqrt{\epsilon}$
takes care of the suppression of the term $-\lambda^{(s)}_c
{\cal{P}}_{tc}$ when compared to $P_c=-\lambda^{(d)}_c
{\cal{P}}_{tc}$ (Eq. (\ref{Pqdef})).

It is convenient to introduce 
\begin{equation}
\label{zzeta}
ze^{i\zeta}\equiv R_b\frac{P_u}{P_c}=
-\,\frac{xe^{i\Delta}-C/T}{de^{i\theta}(1+C/T)},
\end{equation}
so that
\begin{equation}
P'=-\frac{1}{\sqrt{\epsilon}}\,P_c\,(1+\epsilon z e^{i(\zeta+\gamma)}).
\end{equation}
For $C/T \approx 0$ one then expects
\begin{equation}
\label{zzetaforCT0}
ze^{i \zeta}\approx \frac{x}{d}e^{i(\pi+\Delta-\theta)}=(2.08 \pm 0.38) 
e ^{i(-19\pm13)}.
\end{equation}

\subsection{$B^+ \to K ^+\bar{K}^0$}
In the $SU(3)$ symmetric case, the $B^+ \to K^+\bar{K}^0$ amplitude is given by
$P$ of Eq. (\ref{Ppipi}):
\begin{eqnarray}
A(B^+\to K^+\bar{K}^0)=P_{KK}&=&
-\lambda^{(d)}_u{\cal{P}}_{tu}-\lambda^{(d)}_c{\cal{P}}_{tc}\nonumber\\
&=&P_c\,\left(1- R_b\frac{P_u}{P_c}e^{i \gamma}\right)\nonumber\\
\label{KpK0interfere}
&=&P_c\,\left(1-z e^{i(\zeta+\gamma)}\right).
\end{eqnarray}
We know, however, that $SU(3)$ is broken. 
The main difference between the $B^+ \to \pi^+ K^0$ and $B^+ \to K^+ \bar{K}^0$
amplitudes stems then from the fact that in the first decay the $q\bar{q}$ pair 
created after the weak decay is composed of light quarks, while in the other decay
it is an $s\bar{s}$ pair.
Now, it is known that processes, in which such newly produced 
quarks $q$ and $\bar{q}$
end up in different (and separated by large rapidity gap) hadrons, are
suppressed for strange quarks more than for the light ones.
Let us therefore write 
 the $B^+ \to K^+ \bar{K}^0$ amplitude for the case of $SU(3)$
breaking (parametrized by $\kappa <1$) as
\begin{eqnarray} 
\label{PKK}
P_{KK}=\kappa P_c \left(1-z e^{i(\zeta +\gamma)}\right).
\end{eqnarray}

\subsection{Estimate of $SU(3)$ breaking}
\label{estimate}
\begin{figure}
\caption{B decays to $\pi K$ and $K\bar{K}$ final states. Penguin $\bar{b} \to
\bar{s}$ transitions are
denoted by crosses. Vertical dashed line
symbolizes onset of long range dynamics.
Loops symbolize transition through many-body intermediate states.}
\label{fig1}
\begin{center}
\setlength{\unitlength}{0.7pt}
  \begin{picture}(460,220)
  \put(0,-20){\begin{picture}(200,220)
  \put(10,50){\vector(1,0){85}}
  \put(180,50){\line(-1,0){85}}
  \put(180,80){\vector(-1,0){60}}
  \multiput(58,40)(0,20){10}{\line(0,1){10}}
  \put(110,90){\line(0,1){80}}
  
  \put(10,210){\line(1,0){85}}
  \put(10,209.5){\line(1,0){85}}
  \put(10,210.5){\line(1,0){85}}
  \put(10,209){\line(1,0){35}}
  \put(10,211){\line(1,0){35}}
  \put(10,208.5){\line(1,0){35}}
  
  \put(35,200){\line(1,1){20}}
  \put(35,200.5){\line(1,1){20}}
  \put(35,199.5){\line(1,1){20}}
  \put(35,201){\line(1,1){20}}
  
  \put(35,220){\line(1,-1){20}}
  \put(35,220.5){\line(1,-1){20}}
  \put(35,219.5){\line(1,-1){20}}
  \put(35,219){\line(1,-1){20}}
  
  \put(180,210){\vector(-1,0){85}}
  \put(180,210.5){\line(-1,0){85}}
  \put(180,209.5){\line(-1,0){85}}
  \put(120,180){\line(1,0){60}}
 \put(80,150){\oval(25,25)}
 \put(80,110){\oval(25,25)}
  \put(120,90){\oval(20,20)[bl]}
  \put(120,170){\oval(20,20)[tl]}
  \put(185,187){\makebox{\large {${K^0}$}}}
  \put(185,60){\makebox{\large {${\pi^+}$}}}
  \put(150,193){\makebox{\large {$\bar{s}$}}}
  \put(15,188){\makebox{\large {$\bar{b}$}}}
  \put(150,165){\makebox{\large {${d}$}}}
  \put(150,85){\makebox{\large {$\bar{d}$}}}
  \put(150,55){\makebox{\large {${u}$}}}
  \put(15,55){\makebox{\large {${u}$}}}
  \put(0,120){\makebox{\large {${B^+}$}}}
  \end{picture}}
  
  \put(260,-20){\begin{picture}(200,220)
  \put(10,50){\vector(1,0){85}}
  \put(180,50){\line(-1,0){85}}
  \put(180,80){\vector(-1,0){60}}
  \put(180,80.5){\line(-1,0){60}}
  \put(180,79.5){\line(-1,0){60}}
  
  \put(10,208.5){\line(1,0){35}}
  \put(10,209){\line(1,0){35}}
  \put(10,210){\line(1,0){35}}
  \put(10,211){\line(1,0){35}}
  \put(10,209.5){\line(1,0){35}}
  \put(10,210.5){\line(1,0){35}}
  
  \put(35,200){\line(1,1){20}}
  \put(35,200.5){\line(1,1){20}}
  \put(35,199.5){\line(1,1){20}}
  \put(35,201){\line(1,1){20}}
  
  \put(35,220){\line(1,-1){20}}
  \put(35,220.5){\line(1,-1){20}}
  \put(35,219.5){\line(1,-1){20}}
  \put(35,219){\line(1,-1){20}}

   \multiput(58,40)(0,20){10}{\line(0,1){10}}
  \put(110,90){\line(0,1){80}}
  \put(110.5,90){\line(0,1){80}}
  \put(109.5,90){\line(0,1){80}}
  \put(10,210){\line(1,0){85}}
  \put(180,210){\vector(-1,0){85}}
  \put(120,180){\line(1,0){60}}
  \put(120,179.5){\line(1,0){60}}
  \put(120,180.5){\line(1,0){60}}
  \put(80,150){\oval(25,25)}
 \put(80,110){\oval(25,25)}
  \put(120,90){\oval(20.5,20.5)[bl]}
  \put(120,90){\oval(19,19)[bl]}
  \put(120,90){\oval(20,20)[bl]}
  \put(120,90){\oval(21,21)[bl]}
  \put(120,170){\oval(20,20)[tl]}
  \put(120,170){\oval(19,19)[tl]}
  \put(120,170){\oval(20.5,20.5)[tl]}
  \put(120,170){\oval(21,21)[tl]}
  \put(185,187){\makebox{\large {$\bar{K}^0$}}}
  \put(185,60){\makebox{\large {${K^+}$}}}
  \put(150,165){\makebox{\large {${s}$}}}
  \put(150,85){\makebox{\large {$\bar{s}$}}}
  \put(15,188){\makebox{\large {$\bar{b}$}}}
  \put(150,190){\makebox{\large {$\bar{d}$}}}
  \put(150,55){\makebox{\large {${u}$}}}
  \put(15,55){\makebox{\large {${u}$}}}
  \put(0,120){\makebox{\large {${B^+}$}}}
  \end{picture}}
\end{picture}
\end{center}
\end{figure}
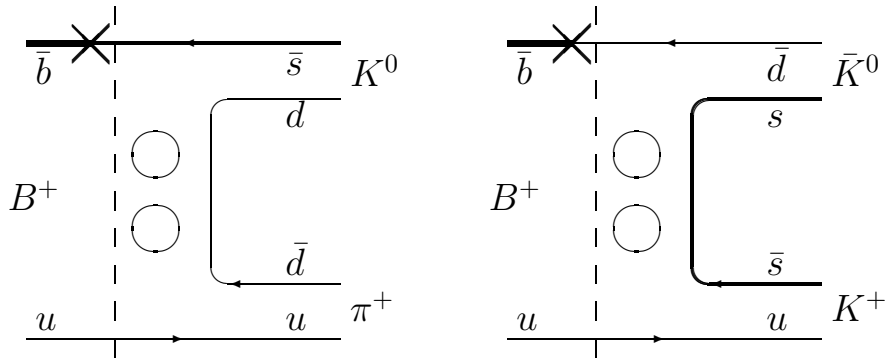
In order to estimate the size of $\kappa$ consider quark line diagrams
corresponding to decays $B^+ \to K^+\bar{K}^0,\pi^+K^0$ as visualized in 
Fig. \ref{fig1}. The decay process starts from the short-distance penguin
transition $\bar{b} \to \bar{s}$ (or $\bar{b} \to \bar{d}$), in Fig. \ref{fig1} denoted by crosses. 
As the highly energetic $\bar{s}$ or $\bar{d}$ quark recedes from the spectator
$u$ quark, a complicated hadronization process sets in.
 The additional $q\bar{q}$ pair observed in the
final state emerges only at the very end of this soft process. 
The complicated nature of the latter is
 visualized in the diagrams of Fig. \ref{fig1} with the help of closed
quark loops, which symbolize the fact 
that the creation of the final $q\bar{q}$ pair may in general go
through various many-body intermediate states.
Thus, the diagrams depict a general situation with {\it all} final state
interactions included. The difference between
the two soft processes shown in Fig. \ref{fig1} is induced by the difference
 in the flavour of the produced
$q\bar{q}$ pair only.
The question is: can we establish the relative size of these two complicated
processes
from the {\it experimental} knowledge of other processes? This can be done if we can
find other, {\it purely soft}, processes 
(i.e. not involving $B$ decays at all), which differ just like
the two processes shown in Fig. \ref{fig1}, 
and on which experimental data exist.

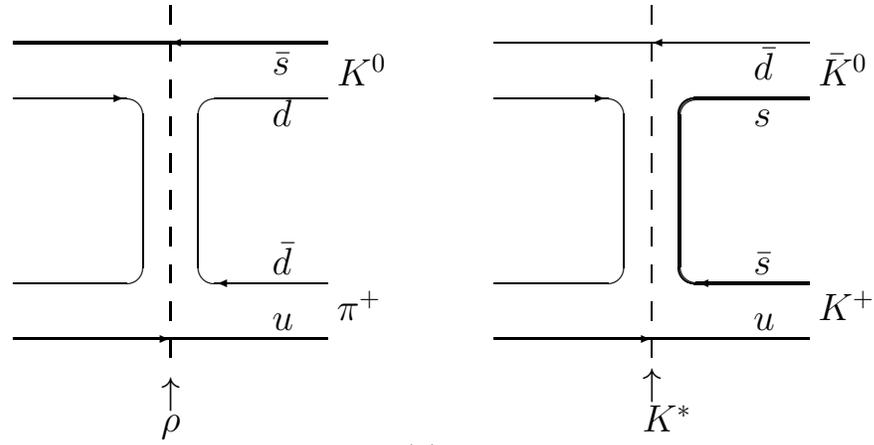
\begin{figure}
\caption{$(a)$ Reggeon exchanges leading to $\pi K$ and $K\bar{K}$ final states
$(b)$ Corresponding product structure of initial and final soft hadron 
interactions. Loops symbolize some of complicated 
intermediate hadronic states not shown in upper diagrams explicitly}
\label{fig2}
\begin{center}
\setlength{\unitlength}{0.7pt}
  \begin{picture}(460,540)
  
  \put(0,300){\begin{picture}(460,240)
  \put(0,20){\begin{picture}(200,220)
  \put(10,50){\vector(1,0){85}}
  \put(180,50){\line(-1,0){85}}
  \put(10,80){\line(1,0){60}}
  \put(180,80){\vector(-1,0){60}}
  \put(80,90){\line(0,1){80}}
  \multiput(95,40)(0,20){10}{\line(0,1){10}}
  \put(110,90){\line(0,1){80}}
  \put(10,210){\line(1,0){85}}
  \put(10,209.5){\line(1,0){85}}
  \put(10,210.5){\line(1,0){85}}
  \put(180,210){\vector(-1,0){85}}
  \put(180,210.5){\line(-1,0){85}}
  \put(180,209.5){\line(-1,0){85}}
  \put(10,180){\vector(1,0){60}}
  \put(120,180){\line(1,0){60}}
  \put(70,90){\oval(20,20)[br]}
  \put(70,170){\oval(20,20)[tr]}
  \put(120,90){\oval(20,20)[bl]}
  \put(120,170){\oval(20,20)[tl]}
  \put(185,187){\makebox{\large {${K^0}$}}}
  \put(185,60){\makebox{\large {${\pi^+}$}}}
  \put(90,15){\makebox{\large {$\uparrow$}}}
  \put(90,0){\makebox{\large {$\rho$}}}
  \put(150,193){\makebox{\large {$\bar{s}$}}}
  \put(150,165){\makebox{\large {${d}$}}}
  \put(150,85){\makebox{\large {$\bar{d}$}}}
  \put(150,55){\makebox{\large {${u}$}}}
  \end{picture}}
  \put(260,20){\begin{picture}(200,220)
  \put(10,50){\vector(1,0){85}}
  \put(180,50){\line(-1,0){85}}
  \put(10,80){\line(1,0){60}}
  \put(180,80){\vector(-1,0){60}}
  \put(180,80.5){\line(-1,0){60}}
  \put(180,79.5){\line(-1,0){60}}
  \put(80,90){\line(0,1){80}}
   \multiput(95,40)(0,20){10}{\line(0,1){10}}
  \put(110,90){\line(0,1){80}}
  \put(110.5,90){\line(0,1){80}}
  \put(109.5,90){\line(0,1){80}}
  \put(10,210){\line(1,0){85}}
  \put(180,210){\vector(-1,0){85}}
  \put(10,180){\vector(1,0){60}}
  \put(120,180){\line(1,0){60}}
  \put(120,179.5){\line(1,0){60}}
  \put(120,180.5){\line(1,0){60}}
  \put(70,90){\oval(20,20)[br]}
  \put(70,170){\oval(20,20)[tr]}
  \put(120,90){\oval(20.5,20.5)[bl]}
  \put(120,90){\oval(19,19)[bl]}
  \put(120,90){\oval(20,20)[bl]}
  \put(120,90){\oval(21,21)[bl]}
  \put(120,170){\oval(20,20)[tl]}
  \put(120,170){\oval(19,19)[tl]}
  \put(120,170){\oval(20.5,20.5)[tl]}
  \put(120,170){\oval(21,21)[tl]}
  \put(185,187){\makebox{\large {$\bar{K}^0$}}}
  \put(185,60){\makebox{\large {${K^+}$}}}
  \put(150,165){\makebox{\large {${s}$}}}
  \put(150,85){\makebox{\large {$\bar{s}$}}}
  \put(150,190){\makebox{\large {$\bar{d}$}}}
  \put(150,55){\makebox{\large {${u}$}}}
  \put(90,20){\makebox{\large {$\uparrow$}}}
  \put(90,0){\makebox{\large {$K^*$}}}
  \end{picture}}
  \put(220,0){\makebox{$(a)$}}
\end{picture}}

\put(0,10){\begin{picture}(460,240)
  \put(0,10){\begin{picture}(200,220)
  \put(10,50){\vector(1,0){75}}
  \put(180,50){\line(-1,0){75}}
  \put(10,80){\line(1,0){30}}
  \put(180,80){\vector(-1,0){30}}
  \put(50,90){\line(0,1){80}}
  \multiput(95,40)(0,20){10}{\line(0,1){10}}
  \put(140,90){\line(0,1){80}}
  \put(10,210){\line(1,0){75}}
  \put(10,209.5){\line(1,0){75}}
  \put(10,210.5){\line(1,0){75}}
  \put(180,210){\vector(-1,0){75}}
  \put(180,210.5){\line(-1,0){75}}
  \put(180,209.5){\line(-1,0){75}}
  \put(10,180){\vector(1,0){30}}
  \put(150,180){\line(1,0){30}}
  \put(40,90){\oval(20,20)[br]}
  \put(40,170){\oval(20,20)[tr]}
  \put(150,90){\oval(20,20)[bl]}
  \put(150,170){\oval(20,20)[tl]}
  
  \put(115,100){\oval(25,25)}
  \put(115,140){\oval(25,25)}
    
   \put(10,125){\makebox{\large {${u,d}$}}}
  \put(150,193){\makebox{\large {$\bar{s}$}}}
  \put(150,125){\makebox{\large {${d}$}}}
  \put(150,55){\makebox{\large {${u}$}}}
  \end{picture}}
  \put(260,10){\begin{picture}(200,220)
  \put(10,50){\vector(1,0){75}}
  \put(180,50){\line(-1,0){75}}
  \put(10,80){\line(1,0){30}}
  \put(180,80){\vector(-1,0){30}}
  \put(180,80.5){\line(-1,0){30}}
  \put(180,79.5){\line(-1,0){30}}
  \put(50,90){\line(0,1){80}}
   \multiput(95,40)(0,20){10}{\line(0,1){10}}
  \put(140,90){\line(0,1){80}}
  \put(140.5,90){\line(0,1){80}}
  \put(139.5,90){\line(0,1){80}}
  \put(10,210){\line(1,0){75}}
  \put(180,210){\vector(-1,0){75}}
  \put(10,180){\vector(1,0){30}}
  \put(150,180){\line(1,0){30}}
  \put(150,179.5){\line(1,0){30}}
  \put(150,180.5){\line(1,0){30}}
  \put(40,90){\oval(20,20)[br]}
  \put(40,170){\oval(20,20)[tr]}
  \put(150,90){\oval(20.5,20.5)[bl]}
  \put(150,90){\oval(19,19)[bl]}
  \put(150,90){\oval(20,20)[bl]}
  \put(150,90){\oval(21,21)[bl]}
  \put(150,170){\oval(20,20)[tl]}
  \put(150,170){\oval(19,19)[tl]}
  \put(150,170){\oval(20.5,20.5)[tl]}
  \put(150,170){\oval(21,21)[tl]}
   
  \put(115,100){\oval(25,25)}
  \put(115,140){\oval(25,25)}

  \put(150,125){\makebox{\large {${s}$}}}
  \put(10,125){\makebox{\large {${u,d}$}}}
  \put(150,190){\makebox{\large {$\bar{d}$}}}
  \put(150,55){\makebox{\large {${u}$}}}
  \end{picture}}
  \put(220,0){\makebox{$(b)$}}
\end{picture}}

\end{picture}
\end{center}
\end{figure}

Imagine now that the $\bar{s}u$ or $\bar{d}u$ 
states - from which the soft hadronization processes 
of Fig.
\ref{fig1} start - are generated not via a $B$-decay, but
by a different initial process, nearly identical for both $\bar{q}u$
states,
as shown in Fig. \ref{fig2} (b) (the following ideas lie at the foundations of
the Regge-based estimates of strong decay widths as performed in ref. \cite{Kwiecinski}).
This is a process in which two mesons collide and form an intermediate
$\bar{s}u$ or $\bar{d}u$ state. 
The initial process, leading to the $\bar{s}u$ or $\bar{d}u$ state,
and the final process, leading to $K^0\pi^+$ or $\bar{K}^0K^+$, may be redrawn
together as shown in Fig. \ref{fig2} (a). In this figure, the complicated nature of soft
interactions, visualized in Fig. \ref{fig2} (b) by quark loops, is not shown
 explicitly at all. Obviously, however,
if we extract {\it from experiment} the
amplitudes corresponding to the topology of the 
diagrams shown in Fig. \ref{fig2} (a), the effect of
 {\it all} such soft
interactions will be included in our extracted amplitudes.
At energy $s=m^2_B$, which is relevant for our case, 
these amplitudes are dominated
by $\rho$ and $K^*$ Regge exchanges in the $t$-channel, 
and the experimental amplitudes may be expressed in terms of the corresponding
Regge parameters. 
Since these parameters are extracted directly from high
energy scattering experiments, 
their values take into account  
{\it all} final state interactions in the $s$-channel, even those
generated by Pomeron exchange (the so-called Reggeon-Pomeron cuts) - the only strong interaction allowed 
after the  final $d\bar{d}$ or $s\bar{s}$ pair is produced.

The experimental amplitudes corresponding to Fig. \ref{fig2} (a)
may be parameterized in terms of the product of Regge couplings and Regge
propagators. Thus, we have (for the left and right diagrams, respectively):
\begin{eqnarray}
&g_{\rho KK}g_{\rho\pi\pi}\left(s/s_0\right)^{\alpha_{\rho}(t)}&\\
&g_{K^* \pi K}g_{K^*\pi K}\left(s/s_0\right)^{\alpha_{K^*}(t)}&
\end{eqnarray}
where $\alpha_M(t)$ is the Regge trajectory for meson $M$,
 given in terms of its intercept $\alpha_0(M)$ and the universal
slope $\alpha'$ by:
\begin{eqnarray}
\label{Reggetraj}
\alpha_M(t)&=&\alpha_0(M)+\alpha't
\end{eqnarray}
with
$s_0 =(\alpha')^{-1} \approx 1~GeV^2$ being the scale parameter
relevant for soft processes. When one takes into account that
the intercepts $\alpha_0(M)$ are determined in soft processes,
consistent application of Eq. (\ref{Reggetraj}) requires
 that one cannot use any other value for the scale parameter
(like e.g. $s_0=m^2_B$). The scale $s_0=1~GeV^2$ 
is fixed by the Regge behaviour 
as experimentally observed in soft processes. 
It is irrelevant here that the true Regge behaviour sets in at an energy much higher than
$s_0$ 
(in fact the Regge formula describes also the region of low energies, albeit
only in an average way).

Let us now discuss the issue of $SU(3)$ symmetry breaking.
Regge amplitudes result from the summation over exchanged resonances
(the sums being performed either in the $s$- or in the $t$- channel).
In principle, the couplings of external $\pi$, $K$ mesons to these individual resonances 
may break $SU(3)$ . 
One may then wonder 
if such 
$SU(3)$ breaking effects could not add up in the summation procedure
in an uncontrolled way
and lead to unknown $SU(3)$ breaking in Reggeon couplings.
Yet, please note that in fact we are talking not about the calculation of $SU(3)$
properties of Regge amplitudes from those of the resonances, but about the parametrization of 
{\it experimental} flavour-exchange amplitudes at such energies
at which Regge behaviour is observed. It is the experimentally observed
energy dependence of these amplitudes
as well as  their {\it absolute} and {\it relative} magnitudes 
that determine  Regge
parametrization. These things are known from the fits to the cross-sections' 
data: the observed energy dependence
fixes the intercepts, while their absolute size fixes Reggeon
couplings.
In fact, from the relative magnitude of the amplitudes
it is known that the extracted couplings of the leading non-Pomeron Reggeons
to the
external particles (i.e. $\pi, K$) satisfy $SU(3)$ symmetry well
 \cite{Irving} (see also \cite{Barger}).
 An analogous statement is true for various hadronic couplings
 (c.f. the successes of $SU(3)$-symmetric parametrization of the $MBB'$ couplings
 of ground-state mesons and baryons in
 terms of $SU(3)$ parameters $F$ and $D$, and many other similar examples). In
 fact,
  $SU(3)$ breaking in hadronic couplings at low energies is much weaker than in
  hadron masses. A corresponding statement in Regge phenomenology is that
  $SU(3)$ breaking in Regge residues is much less important than in the
  intercepts.
  In the following we shall therefore accept $SU(3)$ of Reggeon couplings
which means that
$g_{\rho KK}g_{\rho\pi\pi} = g_{K^* \pi K}g_{K^*\pi K} $.
\footnote{Although $SU(3)$ symmetry is not satisfied by Pomeron couplings, 
this does not affect our estimates at all, as our scheme does not use
these couplings, but only those
of the non-leading Reggeons, as required by the topology of the diagrams under
consideration
and extracted from experiments relevant for these topologies.}

At $s=m^2_B$ 
the size of the 
$K^*$-exchange Regge amplitude relative to that of the
$\rho$ exchange is then clearly given by
\begin{eqnarray}
\label{softKstartorho}
&(m^2_B/s_0)^{\alpha_0(K^*)-\alpha_0(\rho)}.&
\end{eqnarray}

The fact that the initial stage of the collision process, leading to the
intermediate $\bar{s}u$ or $\bar{d}u$ state, is the same in both cases
(see Fig. \ref{fig2} (b)) means that Eq. (\ref{softKstartorho}) provides also
a good estimate of $\kappa$ in $B$ meson decays. The irrelevance of the initial
process leading to $\bar{s}u$ or $\bar{d}u$ 
may be seen in a yet another way. Namely,
 the intercepts of
 the leading Regge trajectories depend on the flavour of the 
 exchanged quarks, and it is known that this dependence is approximately
 additive, i.e.
 \begin{eqnarray}
 \alpha_0(\rho)&=&\alpha_{0n}+\alpha_{0n},\\
 \alpha_0(K^*)&=&\alpha_{0n}+\alpha_{0s},\\
 \alpha_0(\phi)&=&\alpha_{0s}+\alpha_{0s}
 \end{eqnarray}
 where subscripts $n,s$ correspond to nonstrange and strange quarks.
 The difference $\alpha_0(K^*)-\alpha_0(\rho))=\alpha_{0s}-\alpha_{0n}$
 originates from the difference in the $\bar{q}u$ decay phase only. 
 The dependence on the factor describing the production phase cancels out.
 With
 $\alpha_{0s} < \alpha_{0n}$ we obtain suppression
 of strange quark exchange in the decay phase 
 relative to that of the nonstrange quark.

In conclusion, 
 the ratio of the two amplitudes in Fig. \ref{fig1}
is given by
\begin{equation}
\kappa = (m^2_B/s_0)^{\alpha_0(K^*)-\alpha_0(\rho)}.
\end{equation}
Obviously, even though in the above formula there appears
an expression resembling
 the $K^*$ and $\rho$ Reggeon propagators, no
 $K^*$ or $\rho$ mesons are actually exchanged.
 The formula simply provides an estimate of the relative size of
 effective quark exchanges, all
 soft interactions included.
 
In refs \cite{Irving,Martin} it was  estimated that 
\begin{eqnarray}
\alpha_0(K^*)-\alpha_0(\rho)&\approx &~-0.20.
\end{eqnarray}
Thus, at $m^2_B=27.9~GeV^2$
 one expects
 \begin{equation}
 \kappa \approx 0.50.
 \end{equation}
If one accepts $\alpha_0(K^*)-\alpha_0(\rho)=-0.15$ as sometimes used, one gets
$\kappa \approx 0.60$.\\

One may wonder why 
 the above method of estimating $SU(3)$-breaking 
 should be preferred to calculations based on effective field theories 
(and the factorization approach in particular).
 Indeed, it is known that QCD factorization and hadron-level
 S-matrix predictions in general do not lead to the same asymptotics as $m_b \to
 \infty$ (see, e.g. \cite{Donoghue1}). Yet, in ref. \cite{Donoghue1} Donoghue et al.
 give preference to the arguments based on
 S-matrix theory, as stated in their concluding section: `For large $m_b$,
there is hope that one can directly calculate the weak matrix elements through
variants of the factorization hypothesis or by pertubative QCD. Final state
interactions will impose limits on the accuracy of such methods, as no existing
technique includes the efffect of inelastic scattering. There must exist, in every
valid theoretical calculation, a region of the parameter space where the
nonperturbative Regge physics is manifest.' 

While the S-matrix approach imposes theoretical requirements on any
calculation performed at quark level, these requirements may or may not be satisfied
by existing quark-level techniques. For example, 
in \cite{DonPetrov} it was argued, in a QCD-based model,
  that  factorization in $B$ decays to two
pseudoscalars holds exactly at the leading order.
These arguments do not apply to our case, however. 
The point is that we have receding colour triplets,
while ref. \cite{DonPetrov} deals with receding colour octets.
While gluon exchanges - as discussed in
\cite{DonPetrov} - may turn octets into singlets, they cannot change 
colour triplets into singlets. The only way to turn colour triplet into a singlet is
through an exchange of a colour triplet, i.e. a quark (not considered in
\cite{DonPetrov}).
Such an exchange necessarily involves flavour exchange as well.
In the abstract of \cite{Donoghue1} we read:
`flavour off-diagonal FSI are suppressed by a power of $m_B$, but are likely to be
significant at $m_b \approx 5~GeV$'.
It is the $SU(3)$ breaking in such flavour exchanges that is estimated in our approach 
with the help of Regge arguments.

\subsection{Constraint from branching ratios}
The CP-averaged branching ratios for the 
$B^+ \to \pi^+K^0, K^+\bar{K}^0$ decays are
given by
\begin{eqnarray}
\langle {\cal{B}}(B^+ \to \pi^+ {K}^0) \rangle_{CP}&\approx &
\frac{1}{\epsilon}|P_c|^2\,(1+2\,\epsilon\, z\, \cos \zeta \cos \gamma),\nonumber\\
\label{BRatios}
\langle {\cal{B}}(B^+ \to K^+ \bar{K}^0) \rangle_{CP}&=& \kappa ^2
|P_c|^2\,(1+z^2-2z \cos \zeta \cos \gamma).
\end{eqnarray} 
Thus, we find that
\begin{eqnarray}
\label{conditionRKKpiK}
R^{KK}_{\pi K}\equiv \frac{\langle {\cal{B}}(B^+ \to K^+ \bar{K}^0) \rangle_{CP}}
{\langle {\cal{B}}(B^+ \to \pi^+ {K}^0) \rangle_{CP}}
&=&\epsilon \,\kappa ^2\,
\frac{1+z^2-2z \cos \zeta \cos \gamma}{1+2\,\epsilon\, z\, 
\cos \zeta \cos \gamma}.
\end{eqnarray}
The experimental branching ratios for $B^+ \to \pi^+ K^0, K^+\bar{K}^0$,
and $B^0\to K^0\bar{K}^0$ decays are (in our approximation the amplitudes for
 $B^+ \to K^+\bar{K}^0$
and $B^0 \to K^0\bar{K}^0$ are identical):
\begin{eqnarray}
\langle {\cal{B}}(B^+ \to \pi^+ {K}^0) \rangle_{CP}&=&23.1 \pm 1.0,\nonumber\\
\langle {\cal{B}}(B^+ \to K^+ \bar{K}^0)
\rangle_{CP}&=&1.36^{+0.29}_{-0.27},\nonumber\\
\langle {\cal{B}}(B_d \to K^0 \bar{K}^0) \rangle_{CP}&=&0.96^{+0.21}_{-0.19}.
\end{eqnarray}
Thus, one finds 
\begin{eqnarray}
R^{KK}_{\pi K}&=&0.059\pm 0.012,
\end{eqnarray}
or 
\begin{eqnarray}
R^{KK}_{\pi K}&=&0.049\pm 0.008,
\end{eqnarray}
where the first (second) line is obtained if 
the $B^+ \to K^+ \bar{K}^0$
(the average of  
the $B^+ \to K^+ \bar{K}^0$ and $B_d \to K^0 \bar{K}^0$) 
branching ratio(s)
is used.
 For illustration purposes, let us accept $R^{KK}_{\pi K} \approx 0.055$.
 Solving Eq. (\ref{conditionRKKpiK}) for this value of $R^{KK}_{\pi K}$
  yields one positive
solution only, a constraint on $z$ and $\zeta$:
\begin{eqnarray}
\label{zzetaRKKpiK}
z&=& (1+\epsilon r) \cos \gamma \cos \zeta + 
\sqrt{r-1+((1+\epsilon r)\cos \gamma \cos \zeta)^2},
\end{eqnarray}
where
\begin{equation}
\label{rdef}
r=\frac{R^{KK}_{\pi K}}{\epsilon \kappa^2} \approx \left\{ 
\begin{array}{cc}
1.1 &{\rm for~ exact~} SU(3),\\
4.4 &{\rm for~ \kappa=0.50}.
\end{array}
\right.
\end{equation}
The analysis of \cite{SZ2005}
indicates that for small $C/T$ the value of
$\zeta$ is of the order of  $ -15^o$ to $-45^o$ 
(see also Eq.(\ref{zzetaforCT0})).
If, as might be expected,
 the relative strong phase of ${\cal{P}}_{tu}$ 
 with respect to ${\cal{P}}_{tc}$ is indeed that small 
 (say $|\zeta| < 30^o $), one estimates that for exact $SU(3)$
\begin{equation}
z \approx 0.9 - 1,
\end{equation}
while if $SU(3)$ is broken, one gets (for $\kappa =0.50$)
\begin{equation}
z \approx 2.3. 
\end{equation}
In both cases, for larger values of $|\zeta|$ one obtains smaller $z$,
 with the minima (0.1 for exact $SU(3)$, $1.4$ for $\kappa=0.50$)
  achieved for $|\zeta| = 180^o$.
  
 The constraint on $z$ and $\zeta$ 
 (Eqs (\ref{conditionRKKpiK},\ref{zzetaRKKpiK}))
depends on $SU(3)$ breaking and
 is shown in Fig. \ref{fig3} with thick lines 
(for $\kappa =0.5,0.6,1.0$). 
Their positioning depends also somewhat on the size of the
$B^+ \to K^+\bar{K}^0$ experimental branching ratio.
An increase (decrease) in the value of the
latter by $0.10$ corresponds roughly to
a decrease (increase) in the size of $\kappa $ by $0.02$ (cf. Eq.(\ref{rdef})).

\subsection{Constraint from asymmetry}
While the $A_{CP}(B^+\to\pi^+K^0)$ asymmetry should be very small,
its $B^+\to K^+\bar{K}^0$ counterpart may be significant, as 
it results from an interference of two penguin contributions of comparable
sizes (see Eq.(\ref{KpK0interfere})). 
Thus, 
 the analysis of this 
asymmetry, given by
\begin{eqnarray}
\label{asymcond}
A_{CP}(B^+\to K^+\bar{K}^0)&=&-\, \frac{2z \sin \zeta \sin \gamma}
{1+z^2-2z\cos \zeta \cos \gamma},
\end{eqnarray}
may provide us with important information on $z$ and $\zeta$.

\begin{figure}[t]
\caption{Constraints on $C/T$ and $P_u/P_c=ze^{i\zeta}/R_b$.
Thick lines: branching ratio constraint from Eq.(\ref{conditionRKKpiK}) 
for $\kappa=0.5,0.6,1.0$.
Dashed lines: curves of constant asymmetry 
$A_{CP}(B^+\to K^+\bar{K}^0)$ 
equal to $-0.06,0.12,0.29$.
Borders of shadowed areas: $|C/T|=0.1,0.2,0.5$. 
Lines of constant $Arg(C/T)$ are also shown.}
\label{fig3}
\begin{center}
\mbox{\epsfbox{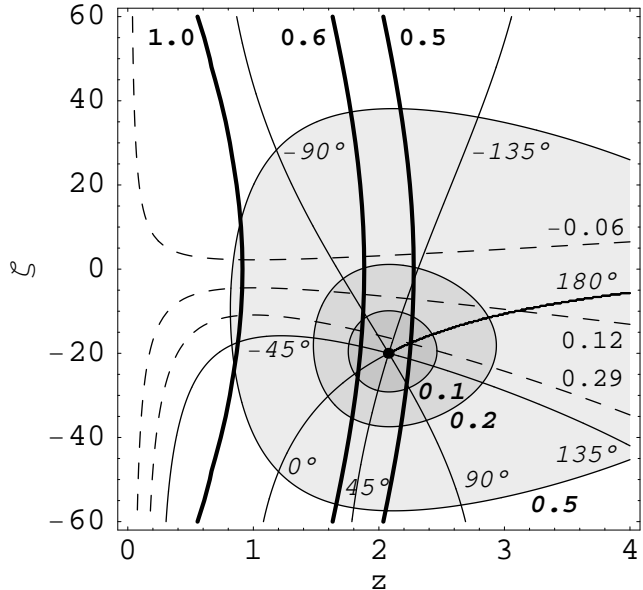}}
\end{center}
 \end{figure}

The experimental data on
 the $A_{CP}(B^+\to K^+\bar{K}^0)$ asymmetry impose another
condition (via Eq.(\ref{asymcond})) on the allowed values of $z$ and $\zeta$. 
The curves along which the asymmetry is constant
(for the experimental values of $0.12^{+0.17}_{-0.18}$)
are shown in Fig. \ref{fig3} as dashed lines. Their positioning is independent of $SU(3)$
breaking.

In the $SU(3)$-breaking case the two constraints cross in the region:
\begin{eqnarray}
z &\approx& 1.8~{\rm to}~2.3 ~~~({\rm for}~\kappa =0.6~{\rm to}~0.5),\nonumber\\
{}\zeta&\approx & -15^o~{\rm to}~0^o.
\end{eqnarray}

\subsection{Implications for $C/T$}
Let us now return to $B \to \pi\pi$. In the following we accept the
central values of $x$, $d$, $\Delta$, and $\theta$.
From Eqs (\ref{dtheta},\ref{xDelta},\ref{zzeta}),  one then 
gets (independently of $\theta $)
\begin{equation}
\label{eqzCT}
z= 2.08 \,\left|\frac{1-0.94 e^{i 59^o}C/T}{1+C/T}\right|.
\end{equation}
For real $C/T$, approaching $z \approx 1$ 
(which for small $\zeta$ is equivalent to the $SU(3)$ case),
 requires large positive values of $C/T$ (around $0.8-1.0$). 
On the other hand, the $SU(3)$ breaking case (for small $\zeta$ corresponding to
$z \approx 1.8-2.3$) clearly needs 
 $C/T$ close to zero. 
 Thus, inclusion of $SU(3)$ breaking in the hadronization stage
is very important.
 
The fit of ref. \cite{Baek2008} does not 
produce large $|C/T|$ in the case of exact $SU(3)$ since
it does not  take into account the condition of Eq. (\ref{conditionRKKpiK}) 
(imposed by the relative size of the
 $B^+ \to \pi^+ K^0$ and $B^+\to K^+\bar{K}^0$ branching ratios), 
 which forces $z
 \le 1$ for any $\zeta$.
Clearly, the fit should be reconsidered with
 the $SU(3)$-breaking expression of Eq. (\ref{PKK}) and
$\kappa \approx 0.5 $.

For any $z$ and $\zeta$, the value of $C/T$ may be
evaluated from Eq. (\ref{zzeta}), provided the $B \to \pi\pi$ parameters are
known sufficiently well. 
For central values of $x$, $d$, $\Delta$, and $\theta$,
the relevant contour plot of $|C/T|$ is presented in Fig. \ref{fig3}, 
with the borders of the
shadowed areas coresponding to $|C/T|=0.1,0.2,0.5$. 
Lines of constant  $Arg(C/T)$ are also shown. 

Fig. \ref{fig3} 
explicitly demonstrates that
for central values of $B\to \pi\pi$ parameters, 
 the data on the $B^+\to K^+\bar{K}^0$ branching ratio and
asymmetry indicate that
 $|C/T|$ is small.
 If present errors in $B\to \pi \pi$ parameters are taken into account,
 the point $C/T=0$ (small central blob in the figure) 
 is shifted by $\Delta z = \pm 0.38$, $\Delta \zeta = \pm 13^o$
 (Eq. (\ref{zzetaforCT0})).
 The qualitative conclusion is not changed.
Further improvement in
the accuracy of the
measurement of the
$B^+\to K^+\bar{K}^0$ and $B\to \pi\pi$ decay parameters
could provide us with more detailed
 information on
$C/T$.

Although our analysis of $B^+ \to \pi^+ K^0, K^+\bar{K}^0$  
does not need or
use the size of $P_c$, the latter
may be estimated from the 
$B^+ \to \pi^+
K^0$ branching ratio. Assuming small $C/T$, 
 one gets
\begin{equation}
P_c=1.03 \pm 0.02,
\end{equation}
an update on the estimate given
in \cite{SZ2005}.

\section{Conclusions}
We have adopted the old Regge model for high-energy soft processes  
in the description
of $SU(3)$ breaking in the hadronization stage of $B$ decays.
We have pointed out that in an analysis of the relative size of $C$, $T$ and 
penguin amplitudes the decay
$B^+ \to \pi^+K^0$ provides important information which has to be taken into
account in addition to that obtained from $B\to \pi\pi$ and $B\to K\bar{K}$.
We have shown
that the data on $B\to \pi\pi$, $\pi^+ K^0$, and $K^+\bar{K}^0$
consistently point to a small value of $|C/T|$.
Further improvement in
the accuracy of the relevant
measurements could tell us more about the actual value of
 $C/T$.

\section{Acknowledgements}
I would like to thank Leonard Le\'sniak for 
providing me with ref. \cite{Irving}.
This work has been partially supported by the Polish Ministry of Science and
Higher Education research project No N N202 248135.

\vfill

\vfill

\end{document}